\documentclass{pasj00}
\draft

\begin{document}
\SetRunningHead{Author(s) in page-head}{Running Head}

\title{Gamma-Ray Burst Polarimeter -- GAP -- aboard 
the Small Solar Power Sail Demonstrator IKAROS }

\author{Daisuke \textsc{Yonetoku},\altaffilmark{1}\email{yonetoku@astro.s.kanazawa-u.ac.jp (DY)}
Toshio \textsc{Murakami},\altaffilmark{1}
Shuichi \textsc{Gunji},\altaffilmark{2}
Tatehiro \textsc{Mihara},\altaffilmark{3}
Tomonori \textsc{Sakashita},\altaffilmark{1}
Yoshiyuki \textsc{Morihara},\altaffilmark{1}
Yukihiro \textsc{Kikuchi},\altaffilmark{1}
Hirofumi \textsc{Fujimoto},\altaffilmark{1}
Noriyuki \textsc{Toukairin},\altaffilmark{2}
Yoshiki \textsc{Kodama},\altaffilmark{1}
Shin \textsc{Kubo},\altaffilmark{4}
and IKAROS Demonstration Team
}
\altaffiltext{1}{Department of Physics, Kanazawa University, 
Kakuma, Kanazawa, Ishikawa 920-1192, Japan}
\email{yonetoku@astro.s.kanazawa-u.ac.jp (DY)}
\altaffiltext{2}{Department of Physics, Faculty of Science, 
Yamagata University, 1-4-12 Koshirakawa Yamagata-city
Yamagata 990-8560, Japan}
\altaffiltext{3}{Cosmic Raditaion Laboratory, RIKEN
2-1 Hirosawa, Wako City, Saitama, 351-0198 Japan}
\altaffiltext{4}{Clear Pulse Co., 
6-25-17, Chuo, Ohta-ku, Tokyo, 143-0024 Japan}
\altaffiltext{5}{
Institute of Space and Astronautical Science (ISAS)
Japan Aerospace Exploration Agency (JAXA)
3-1-1 Yoshinodai, Sagamihara, Kanagawa 229-8510, Japan}

%

\KeyWords{gamma rays: bursts --- gamma rays: polarization
--- gamma rays: observation} 

\maketitle

\begin{abstract}
The small solar power sail demonstrator ``IKAROS'' is 
a Japanese engineering verification spacecraft launched by H-IIA 
rocket on May 21, 2010 at JAXA Tanegashima Space Center.
IKAROS has a huge sail with 20~m in diameter which is made of 
thin polyimide membrane. This sail converts the solar 
radiation-pressure into the propulsion force of IKAROS and 
accelerates the spacecraft. The Gamma-Ray Burst Polarimeter (GAP) 
aboard IKAROS is the first polarimeter to observe the 
gamma-ray polarization of Gamma-Ray Bursts (GRBs) during 
the IKAROS cruising phase. GAP is a tinny detector of 3.8 kg 
in weight and 17~cm in size with an energy range between 50--300~keV.
The GAP detector also plays a role of the interplanetary network (IPN) 
to determine the GRB direction. The detection principle of 
gamma-ray polarization is the anisotropy of the Compton scattering. 
GAP works as the GRB polarimeter with the full coincidence mode 
between the central plastic and the surrounding CsI detectors. 
GAP is the first instrument, devoted for the observation of 
gamma-ray polarization in the astronomical history. 
In this paper, we present the GAP detector and 
its ground and onboard calibrations.
\end{abstract}

\section{Introduction}
 Gamma-Ray Bursts (GRBs) are the most energetic explosion in
the universe and most of all occur at the cosmological distance beyond
the redshift of $z > 1$. The current record for the highest redshift 
is GRB~090423 at $z \sim 8.2$ (e.g. \cite{tanvir2009, salvaterra2009}). 
At the brightest case, the isotropic luminosity reaches 
$10^{54}~{\rm erg~s^{-1}}$. Although many physical property about GRBs 
are revealed after the discovery of afterglows \citep{costa}, 
we have little knowledge about their emission mechanism to produce 
gamma-rays. Theoretically, the prompt emissions and the following 
afterglows are thought to be generated by the synchrotron radiation 
in the relativistic jet. The electrons are accelerated to almost light 
speed by the relativistic shocks, and the strong magnetic field above 
$10^{4}~{\rm Gauss}$ is generated within the jet in a short time 
interval of prompt phase (e.g. \cite{rees1992, piran1999}).
In this case, the polarization degree of 
emitted photons is expected to be very high. Therefore, 
the direct measurement of the polarization degree of the prompt emission 
is a key to probe their emission mechanism and the site.

There are several empirical correlations between the rest-frame 
physical quantities of GRBs and their luminosity 
(or isotropic energy $E_{\rm iso}$). The variability--luminosity 
correlation indicates that more variable events are more luminous 
\citep{fenimore}. The lag--luminosity correlation is reported by 
\citet{norris} and \citet{schaefer01}. Each pulse in the prompt emission 
has a time delay of the soft-band emission compared with the hard-band one.
This correlation shows that the events with large spectral time lag
are dimmer than ones with short lag. These two correlations are
based on the temporal behaviors of prompt emission.
Several correlations concerning the spectral property have also
been suggested. \citet{amati02} found a very tight correlation between 
the spectral peak energy ($E_{p}$) and the isotropic energy ($E_{\rm iso}$) 
(see also \cite{amati06}). This $E_p$--$E_{\rm iso}$ correlation was 
confirmed and extended toward X-ray flashes by \citet{sakamoto04} 
and \citet{lamb04}. Independently, \citet{yonetoku04} reported similar 
but tighter correlation between $E_{p}$ and the 1-second peak 
luminosity ($L_p$) called the $E_p$--$L_p$ correlation.
Moreover, using the jet opening half-angle, 
\citet{ggl04} found that $E_{p}$ strongly correlates with
the collimation-corrected gamma-ray energy ($E_{\gamma}$).
However the physical reasons for these correlations are 
not established yet. These empirical properties may indicate 
that there are well-ordered physics in the complex behaviour 
of prompt GRBs. Then the emission mechanism may play an important 
role in these correlations, and the observation of gamma-ray 
polarization is a key to reveal the emission mechanism.

The first X-ray polarimeter is a Bragg reflection type aboard 
the OSO-8 satellite in 1970s \citep{kestenbaum1976}. 
This polarimeter works for monochromatic X-rays of 2.6~keV and 5.2~keV. 
The first detection of X-ray polarization has been done for Crab nebula
\citep{weisskopf1976} with $15.7 \pm 1.5$~\% at 2.6~keV and 
$18.3 \pm 4.2$~\% at 5.2~keV.
After 30~years of the first Crab observation, \citet{dean2008} measured 
again the polarization of the Crab nebula as $46 \pm 10$~\% in 
the energy range between 15~keV to 8~MeV of SPI and IBIS aboard 
the INTEGRAL satellite. There was no further reports except their one 
but the modulation curve is not shown in their paper.

About the GRB polarization, there were earlier reports of measurement
\citep{coburn, mcglynn}. \citet{coburn} reported a clear detection 
with the polarization degree of $P = 80 \pm 20$~\% in GRB~021206 
using the RHESSI solar observation satellite. 
However, reanalyzing the same data, \citet{rutledge04} failed to ensure 
the results of \citet{coburn}, and concluded that the polarization of 
GRB~021206 is consistent with 0~\%. \citet{wigger04} independently 
checked their analyses and got the result of the polarization degree 
of $41^{+57}_{-44}$~\% which is different from the previous two reports 
but consistent with 0~\%. On the other hand, using INTEGRAL data of 
GRB~041219A, \citet{mcglynn} report the possible detection of the 
polarization with $63^{+31}_{-30}$~\% for a brightest single pulse, 
and also $96^{+39}_{-40}$~\% around the short interval including 
the brightest pulse. 
However they could not show us a modulation curve which is the most
important information for the polarization measurement.
Moreover, they could not estimate a systematic error of modulation 
caused by the complex detector configurations of INTEGRAL SPI and
its coded mask system.
Although these measurements of GRB polarization are 
based on the anisotropy of angular distribution of Compton scattering, 
the detector were not designed to work in the coincidence mode.
Moreover, the ground-based calibrations for the polarized X-ray and 
gamma-ray photons were not performed before their launch, 
so their results are not conclusive and still in debate. 

Recently several types of X-ray and gamma-ray polarimeter have 
been developing. For the Bragg reflection type, 
\citet{kitamoto} develop the X-ray polarimeter with a transmission 
multilayer. They demonstrated that this polarimeter works 
in the energy around 0.1 $\sim$ a few keV.
Using the track of moving electron generated by the photo-electric 
absorption, a polarization of soft X-ray is detected by
\citep{tanimori2004, tamagawa, bellazzini}  in the gas chamber with 
the imaging capability. Based on  this technology of the gas imaging 
polarimeter, a satellite called Gravity and Extreme Magnetism SMEX (GEMS) 
is planned to launch in 2014 \citep{swank}.

However, the typical energy band of GRB is 50--300~keV, 
where the cross-section of the Compton scattering dominates 
that of photo-electric absorption for light element materials. 
Therefore, it is more efficient to use the anisotropic angular 
distribution of the Compton scattered photons. Several types of 
polarization detectors with this process are being developed, 
not only for satellites but also for balloon experiments. 
One example is the proto-type of the PHENEX 
instrument by \citet{gunji1994}. The sensitivity to detect a hard
X-ray polarization is continuously improved 
(see also \cite{gunji1994, gunji1997, kishimoto07}). 
At the balloon flight of Crab observation in 2007, \citet{gunji2007} set 
an upper limit for Crab nebula with the PHENEX instrument.
The Polarization Gamma-ray Observer (PoGO) is a honeycomb shape
detector which is composed with hexagonal well-type scintillator arrays 
work in phoswitch technique \citep{kamae2008}. The octagonal scintillators 
\citep{mihara2004}, and matrix layout using the multi-anode photomultiplier 
tubes \citep{suzuki2006} are also developed. 
These instruments are designed as an instrument with narrow field of view 
to observe steady X-ray and gamma-ray sources, such as the Crab nebula, 
pulsar, black hole candidate, and active galactic nuclei. 
However, for GRB observations, we need a wide field of view. 
\citet{yonetoku2006} showed basic concept of the GRB polarimeter
(GAP) (see also \cite{yonetoku2009, murakami2010}).

In this paper, we shortly introduce the solar power sail satellite 
``IKAROS'', and present the ground-based and also in flight calibration
data to ensure our measurements of polarization of prompt emission of GRBs.

\section{Solar Power Sail -- IKAROS and Design of GRB Polarimeter -- GAP} 
\label{sec:ikaros}

\subsection{The Solar Power Sail : IKAROS}

\begin{figure}
\rotatebox{270}{\includegraphics[width=80mm]{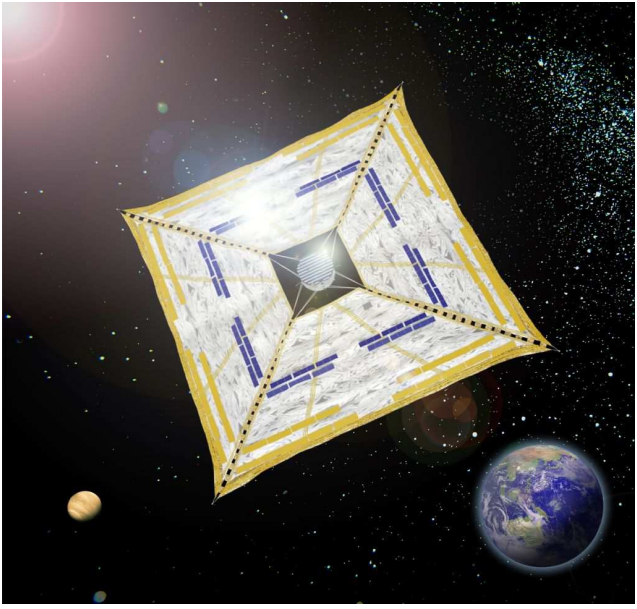}}
\caption{
The schematic view of the IKAROS spacecraft after the sail deployment.
The GAP detector  is mounted at the anti-solar side of the spacecraft. 
Courtesy of JAXA.
}
\label{fig:IKAROS}
\end{figure}

The small solar power sail demonstrator ``IKAROS''  
\citep{kawaguchi2008, mori2009}
is a Japanese engineering verification spacecraft launched 
on May 21, 2010 at JAXA Tanegashima Space Center together with 
the Venus Climate Orbiter AKATSUKI (Planet-C). IKAROS stands for 
``Interplanetary Kite-craft Accelerated by the Radiation Of the Sun''.
The weight of the IKAROS spacecraft is 307~kg, and the size is 
1.58~m in diameter and 0.95~m in height. IKAROS has an extremely thin 
of 7.5~$\mu$m polyimide membrane with 20~m in diameter as shown 
in figure~\ref{fig:IKAROS}. 
Reflecting the photons from the sun, this sail translates solar
radiation pressure to the thrust of the spacecraft. This satellite
rotates with the angular speed of about 1--2~rotation per minute (rpm) 
to keep spreading the sail with the centrifugal forces. 
The purpose of this mission is to demonstrate the solar-sail performance
in the interplanetary space. After the launch, IKAROS successfully 
deployed the solar sail on June 9, 2010, and started sailing to Venus.

According to the current orbit design, IKAROS will pass Venus 
in half a year after the launch and go beyond 2~AU from the earth
in about 2~years. During its cruising phase, 
GAP measures the gamma-ray polarization of the prompt emission of GRBs. 
We also determine the precise position of GRBs using the interplanetary 
network (IPN). In the following sections, we show a design of 
our gamma-ray polarimeter GAP, with its pre-flight and in-flight 
calibration data, and the numerical studies of the Geant~4 simulator.

\subsection{Gamma-ray Burst Polarimeter : GAP}

\begin{figure}
\rotatebox{270}{\includegraphics[width=60mm]{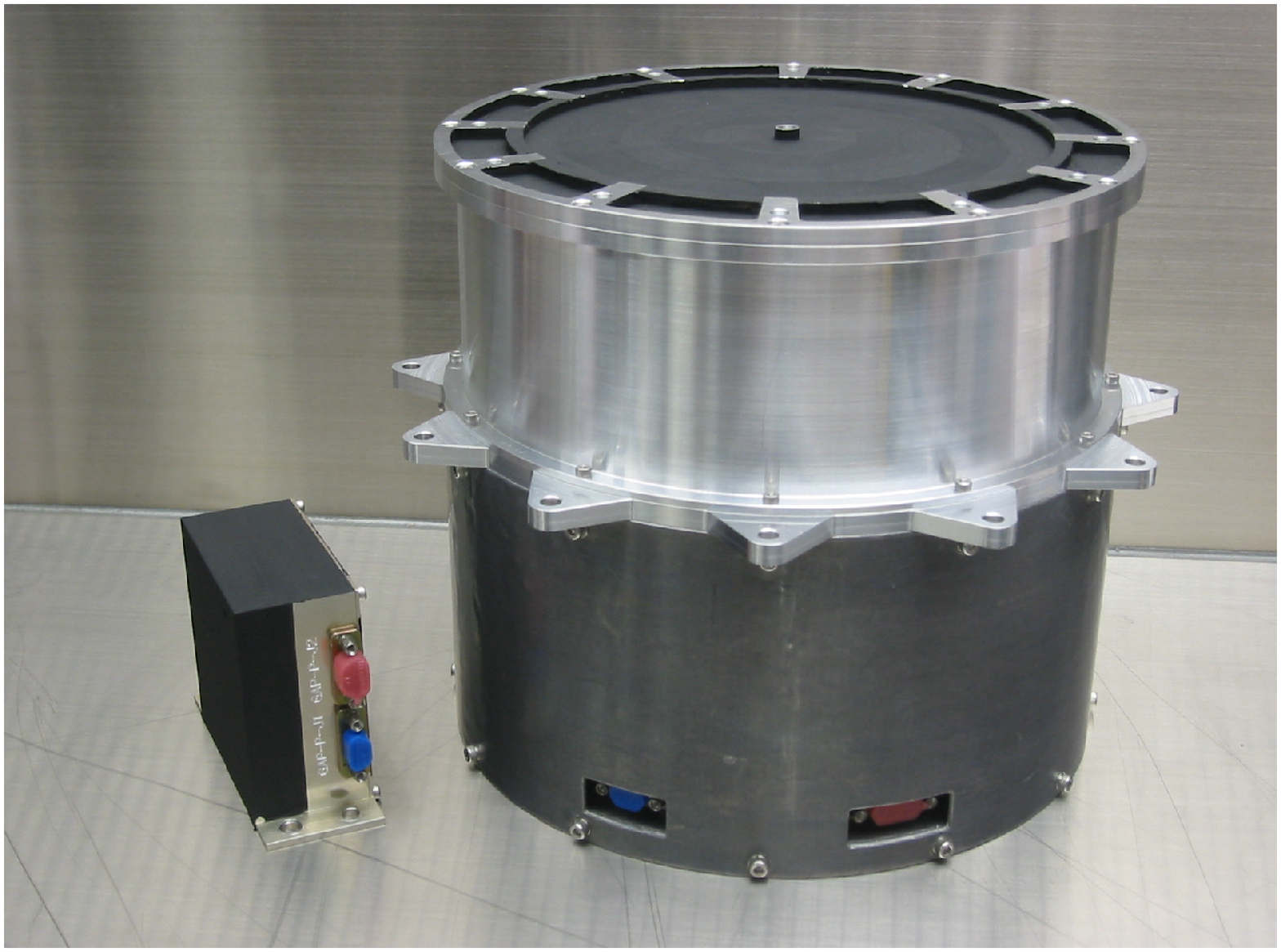}}\\
\rotatebox{270}{\includegraphics[width=60mm]{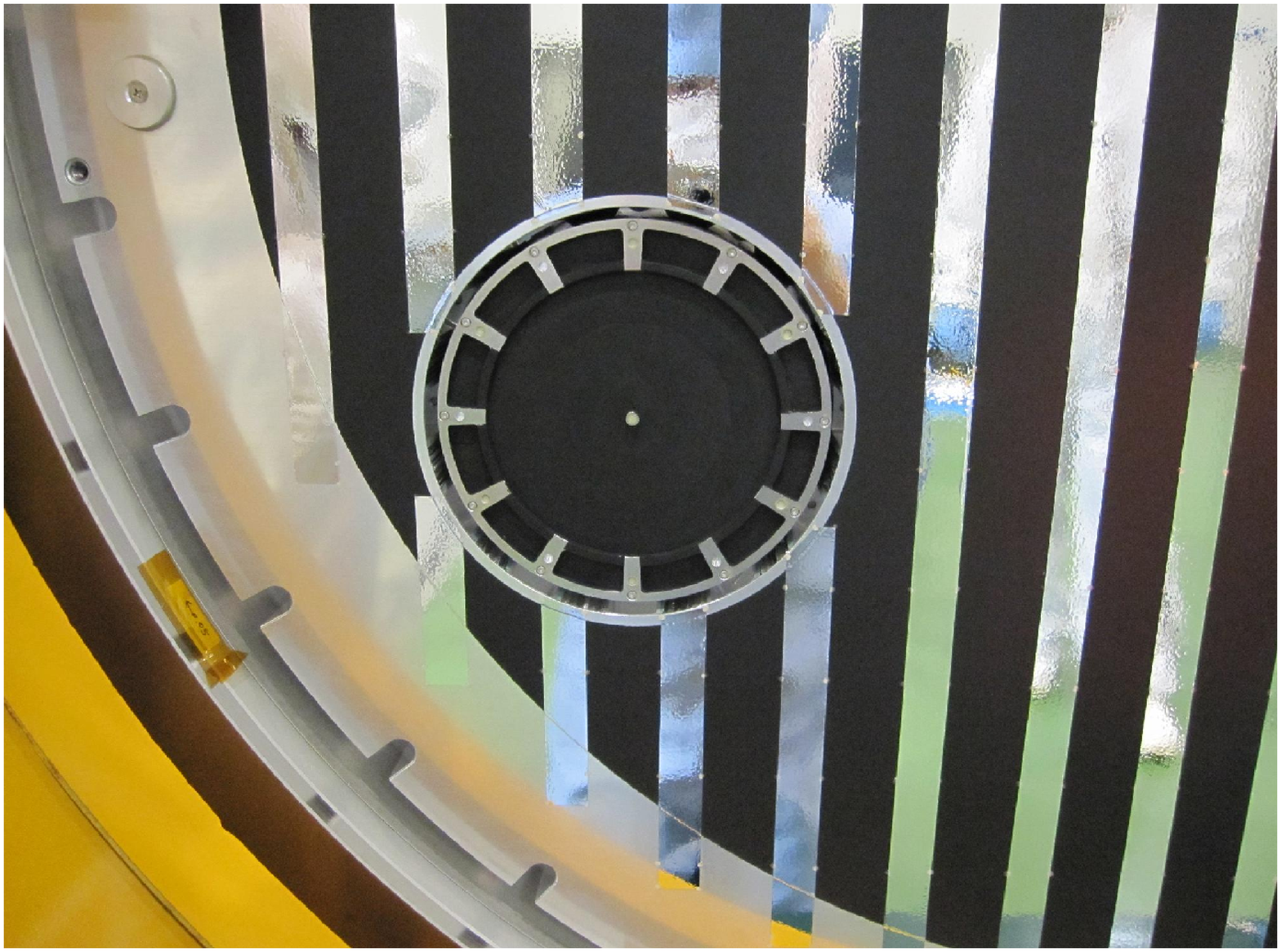}}
\caption{
(upper) Photo of GAP-S and its power unit (GAP-P). 
(lower) Photo of GAP installed on the IKAROS spacecraft. 
A circle in the center of the photo is the top surface
of GAP-S. The strips painted black and white are the surface of 
heat radiation into the space and this side looks anti-solar
direction.}
\label{fig:GAP}
\end{figure}

GAP is the ``GAmma-ray bursts Polarimeter'', which consists of two
components; the detector unit (GAP-S) and the power supply unit (GAP-P).
GAP-S is a very small detector with 17~cm in diameter and also 
16~cm in height as shown in figure~\ref{fig:GAP}. 
A small square box aside GAP-S is GAP-P which contains 
two DC-DC power converters for the electronics and 
two high-voltage modules with a function of current limiter. 
The weight of GAP-S and GAP-P is only 3.69~kg and 0.16~kg, respectively. 
The total power is about 5~W including both GAP-S and GAP-P. 
Hereafter we simply call this system ``GAP''.

GAP is mounted on the bottom panel of IKAROS which always looks 
almost anti-solar direction. Therefore GAP always observe 
the deep universe during the cruising phase of IKAROS. 
The cylindrical detector cases (shassis), except for the detector top, 
are covered by thin lead sheets with 0.3~mm or 0.5~mm thickness 
to avoid the cosmic X- and gamma-ray background, intense soft 
X-ray transients, and solar flare events occurred behind GAP.

\subsection{Design of GAP}

The polarization detection principle of GAP is the angular 
anisotropy of the Compton scattering. The cross section of 
the Compton effect is due to the Klein-Nishina formula;
\begin{eqnarray}
\label{eq:klein-nishina}
\frac{d \sigma}{d \Omega} = 
\frac{r_{0}^{2}}{2} \frac{E^{2}}{E_{0}^{2}}
\Bigl( \frac{E_{0}}{E} + \frac{E}{E_{0}} - 
2 \sin^{2}\theta \cos^{2}\phi \Bigr),
\end{eqnarray}
where, $r_{0}$ is the classical electron radius. 
The $E_{0}$ and $E$ values are the energies of the incident 
and the scattered gamma-ray photon, respectively. 
The $\theta$ denotes the scattering angle, and the $\phi$ 
means the azimuth angle measured from the polarization vector 
(electric field vector) of incident photon.
The relation between $E$ and $E_{0}$ is well known as
\begin{eqnarray}
\label{eq:compton}
E = \frac{E_{0}}{1+\frac{E_{0}}{m_{e}c^2}(1 - \cos \theta)}.
\end{eqnarray}
Here, $m_{e}$ is the electron mass and $c$ is the speed of light 
in a vacuum. Assuming the non-relativistic case by substituting 
$E = E_{0}$, the Klein-Nishina formula is approximately described as
\begin{eqnarray}
\label{eq:thomson}
\frac{d \sigma}{d \Omega} \simeq r_{0}^{2} 
(1 - \sin^{2}\theta \cos^{2} \phi).
\end{eqnarray}
The amplitude of the modulation is basically proportional to 
the function of $\sin^{2} \phi$ in the $\theta = \pi/2$ plane.
By detecting $\sin^{2} \phi$ modulation curve from the astronomical 
phenomena, we can measure the linear polarization degree.

\begin{figure}
\rotatebox{0}{\includegraphics[width=80mm]{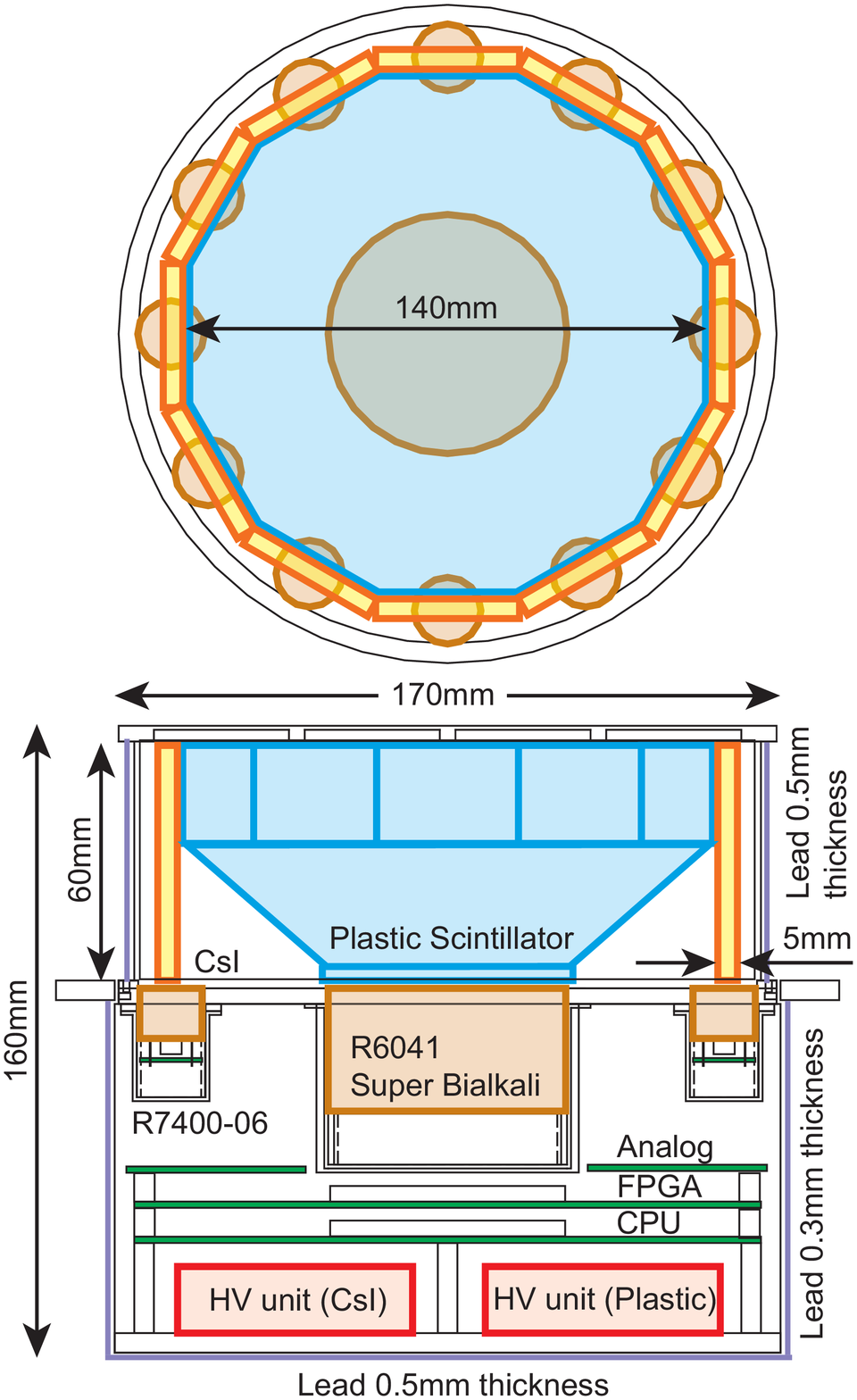}}
\rotatebox{0}{\includegraphics[width=80mm]{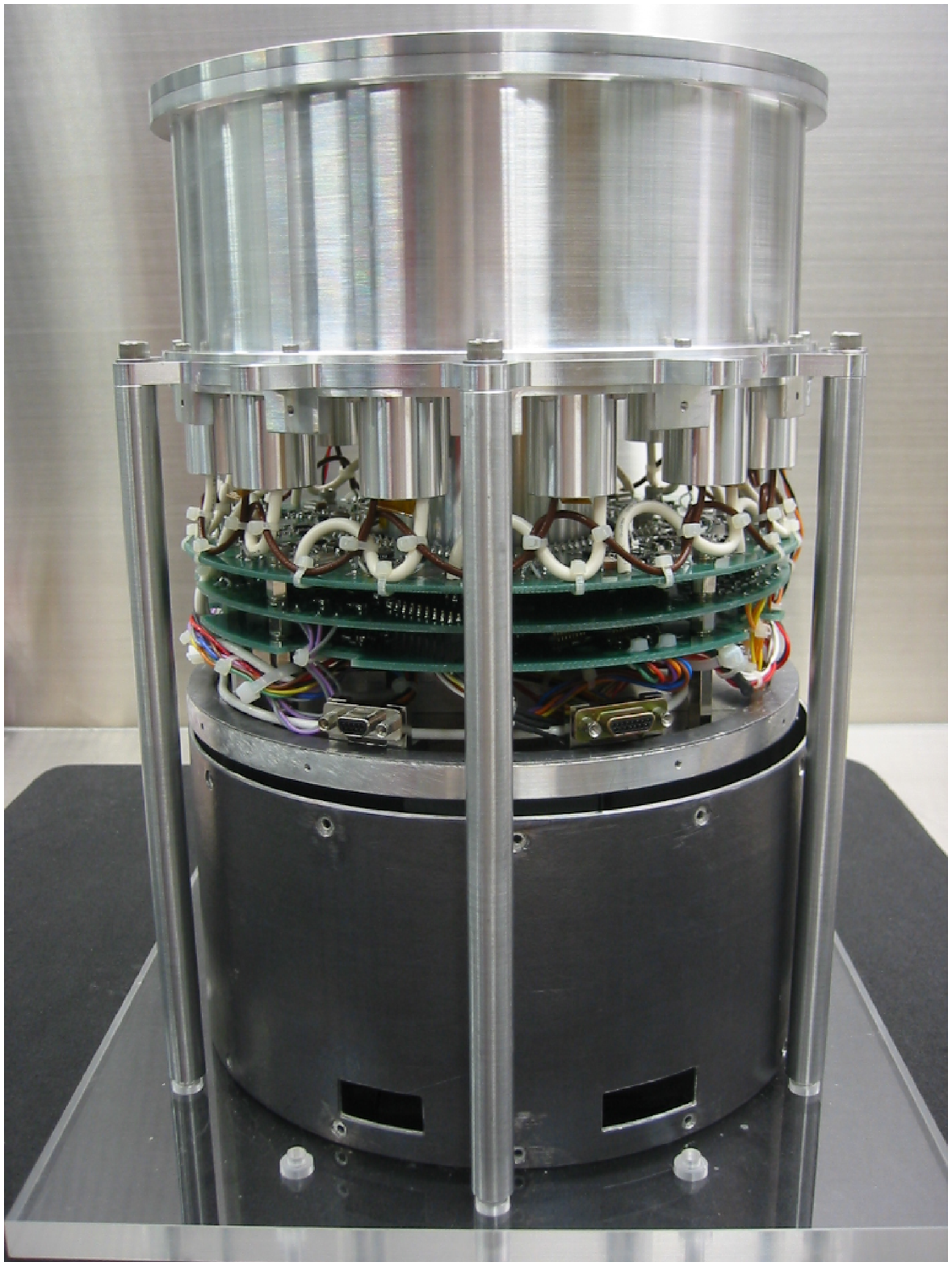}}
\caption{
(left) Schematic view of the GAP sensor. 
A large plastic scintillator with a super bialcari anode of
PMT (R6041) is attached at the center, and 12~CsI scintillators 
with small PMTs (R7400-06) are set around it.
(right) The photo of GAP whose lead shield is opened.
}
\label{fig:model}
\end{figure}

In figure \ref{fig:model}, we show the basic concept of our gamma-ray 
polarimeter GAP. A dodecagon (twelve-sided polygon) plastic 
scintillator with a single non-position sensitive photo-multiplier
tube: PMT (R6041 \cite{hamamatsu}) is attached at the center, 
and 12~CsI scintillators with PMT (R7400-06 \cite{hamamatsu}) 
are set around it. The photo-electric surface of R6041 is 
a super bialcari type, and its electrode structures are re-modeled 
to withstand the strong pyro-shock and vibration during 
the rocket launch. The central plastic works as the Compton scatterer, 
and the angular distribution of scattered photons coinciding with 
the plastic scintillator is measured by the surrounding 
CsI scintillators with the angular resolution of 30~degree each. 
We first considered a multi-anode PMT to know 
a scattering position for the central plastic to have a better
sensitivity for polarization. However, in the developing phase of GAP,
the multi-anode PMTs were known not to survive the strong shock and 
random-vibration environments at the rocket launch. 
Then we avoided the risk and adopted the advantage of high geometrical 
symmetry to reduce the fake modulation. All the PMTs were tested and 
passed the pre-flight qualification vibration levels of JAXA at 
the Industrial Research Institute of Ishikawa. We also passed 
the qualification shock test with 1000~Gsrs level at ISAS/JAXA.

We first designed the depth of the plastic as 7~cm for the pre-flight 
model (PFM) because it fully satisfies one Compton (Thomson) length 
to scatter photons of 100~keV, however to reduce the weight of GAP, 
the depth of the flight model (FM-GAP) is changed to 6~cm. 
Since the scattered photons should escape from the plastic and 
reach CsI, the radius of the plastic is also better to be shorter 
than one Thomson length. The CsI scintillators have to stop 
the scattered photons with high efficiency, we set their thickness 
as 5~mm whose stooping power is almost 100~\% at 100~keV and about 
60~\% at 200~keV.

\subsection{Brief Summary of Electronics and Data Acquisition System}

\begin{figure}
\rotatebox{0}{\includegraphics[width=170mm]{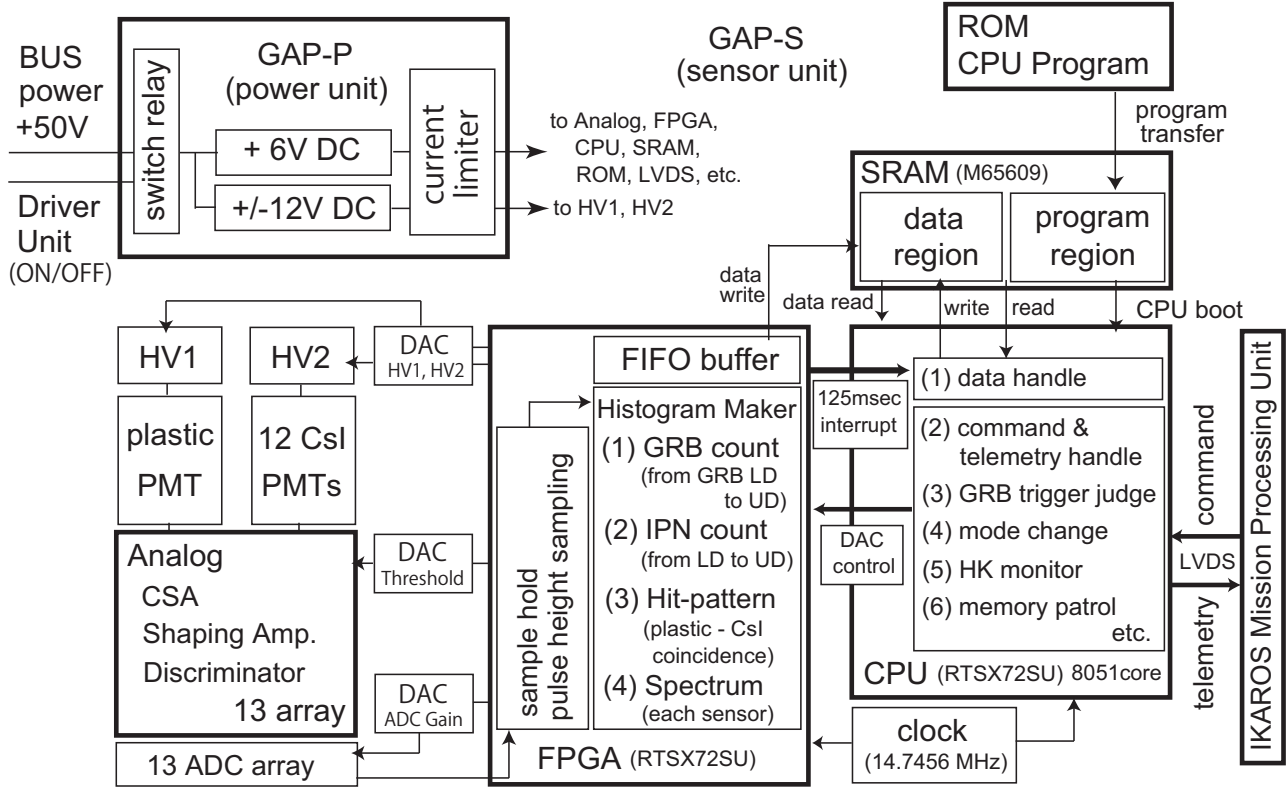}}
\caption{   
Flow chart of the readout circuit. The analog outputs from 13~PMTs are
read out by 13~discrete analog electronics arrays.
Triggered by the output of the discriminator, the FPGA read 
the A/D values of all channels. Then the FPGA analyses the pulse height 
and the makes coincidence within 5~$\mu$sec timing window. 
The CPU appropriately edit the data for each observation mode,
and transfer to the IKAROS mission processing unit and 
mission data recorder.
}
\label{fig:circuit}
\end{figure}

GAP has 13~analog outputs (1 plastic and 12 CsI scintillators).
In figure~\ref{fig:circuit}, we show a flow chart of the readout 
circuit. These 13~signals are read out by 13~discrete analog 
electronics arrays. The standard streams are used in signal readout 
with preamplifiers, shaping amplifiers and discriminators are used.
Once the gamma-ray photon is triggered by any PMT, 
analog signals of all PMTs are converted to digital signals 
with analog-to-digital converter, and the following FPGA reads 
all their digital values. Then, the FPGA analyzes all pulse heights
within 5~$\mu$sec timing window, and selects the plastic--CsI 
coincidence events as the polarization signals.

Every 125~msec, the FPGA interrupts the CPU processing and transfer 
the histogram data created by FPGA to CPU. The CPU edits the data type 
of lightcurve, spectra, and hit-pattern of polarization for each 
observation mode (see below). The edited data by CPU are transfered to 
the IKAROS Mission Processing Unit (MPU), and the data is stored as 
the telemetry packet in the mission data recorder.

The CPU also recognizes the commands and works as the human--GAP 
interface. It controls several Digital-to-Analog Converters 
together with FPGA. We can monitor House Keeping (HK) data 
such as temperature, voltage, current and simple light curves.
We employed a memory patrol system with the majority logic in the CPU, 
so the single-bit memory error can be repaired by the CPU itself. 
This function reduces the probability of CPU hang-up and/or 
miss operations.
GAP used radiation tolerant and MIL class ASICs for the core devices, 
such as FPGA, CPU, SRAM, LVDS but not for ROMs. Almost all key
devices of the GAP circuit cleared two types of test such as 
$^{60}$Co gamma-ray irradiation and 200~MeV of proton irradiation to
select the qualified devices at the Tokyo Metropolitan Isotope
Research Center and at the Wakasa-Wan Energy Research Center.
 
When we switch the GAP power unit on, the CPU program written in 
the ROM device is transfered to the program region of the radiation 
tolerant SRAM. Then the CPU starts to work and the ROM device is 
switched off. The core of CPU is equivalent to the 8051 processor 
which is burned in the same FPGA 
device RTSX72SU. The observation data and the HK data are temporary 
stored in the data region of SRAM, and they are transfered to 
the IKAROS mission processing unit with the LVDS communication 
when the CPU gets the data request commands.
Two high voltage modules are supplied from Meisei Electric Co.,Ltd, 
these are the same type used for the Suzaku and the Kaguya satellites 
of JAXA.

\section{Observation Mode} \label{sec:selection}
The GAP has the calibration mode and 2~observation modes 
(GRB mode and CRAB mode). The onboard CPU is automatically 
monitoring the gamma-ray count rate in every one second. 
If the count rate suddenly increases above the preset value, 
the CPU switches its status to the GRB mode. 
The judge of a GRB trigger is performed with the increase of 
count rate between the energy range of GRB\_LD and UD about 220~keV.
Here, the GRB\_LD is the lower discriminator level for 
the GRB judgement. We can set this GRB\_LD for each sensor, 
and the nominal value is about $60~{\rm keV}$ for 12~CsI 
scintillators and $20~{\rm keV}$ for the plastic scintillator.
We set this GRB\_LD much higher than the sensor LD to avoid 
the noise fluctuations and also to avoid the intense 
soft X-ray transient, mainly solar flares.

\subsection{Calibration Mode and onboard Calibration}
We mounted seven weak radio isotope sources of $^{241}$Am 
on the top panel of GAP, one for two CsI scintillators and
one for the plastic at the center. The count rate of $^{241}$Am 
is about 10~Hz per CsI. In the calibration mode, 
we obtain the detail spectra with 64~energy channels for 
each sensor. Therefore we can measure the pulse height of 
$59.5~{\rm keV}$ line of $^{241}$Am with high precision. 
This spectral data contains only the single-hit photon events
because the coincidence events, e.g. plastic--CsI and CsI--CsI, 
lose some of the energy and it is lower than 59.5~keV.
Since the integration time can be chosen from 30~sec to 4.3~hours 
with 2~minuets steps, we can flexibly count up photons in 
the environment with any background rate, and avoid 
the buffer overflow.

\begin{figure}
\rotatebox{0}{\includegraphics[width=90mm]{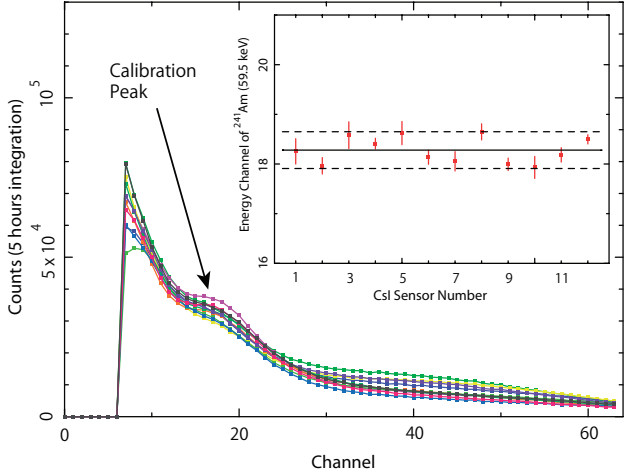}}
\caption{Onboard calibration spectra of 12~CsI scintillators 
with $^{241}$Am (59.5~keV) peak.
All spectra show low-energy cut at around 23~keV with the LD.
The panel inserted on the figure shows the $^{241}$Am 
peak channel of 12~CsI scintillators. The solid line is 
the weighted average and two dashed lines show boundaries of 
2~\% deviations. We succeeded in adjusting all PMT gain within 
2~\% level.
}
\label{fig:calspec}
\end{figure}

In figure~\ref{fig:calspec}, we show 12~CsI spectra obtained
in orbit. This spectra are equivalent to 5~hours integration
(1~hour $\times$ 5) after the careful gain adjustments. 
The 59.5~keV line of $^{241}$Am is clearly observed.
The inserted panel in figure~\ref{fig:calspec} shows the uniformity 
of the peak energy channel of the calibration line for each CsI sensor.
The best fit line is 18.2~channel and two dashed lines show
the upper and lower boundary of 2~\% deviations from the best fit.
We succeeded in adjusting 12~CsI gains within 2~\% level.
We, of course, monitor these gains once in every two weeks,
and confirmed they are quite stable within the 2~\% deviations 
during the first two months.

\subsection{GRB Mode}
Once the CPU switches to the GRB mode, CPU records the coincidence 
hit-pattern (polarization data), lightcurve, and spectrum during 
176~seconds since the time of GRB trigger ($T = 0$). 
Then, the GAP's ring buffer memory is frozen, and GAP-CPU also records 
the data of 16~seconds before the trigger ($-16 \le T < 0$).

\begin{figure}
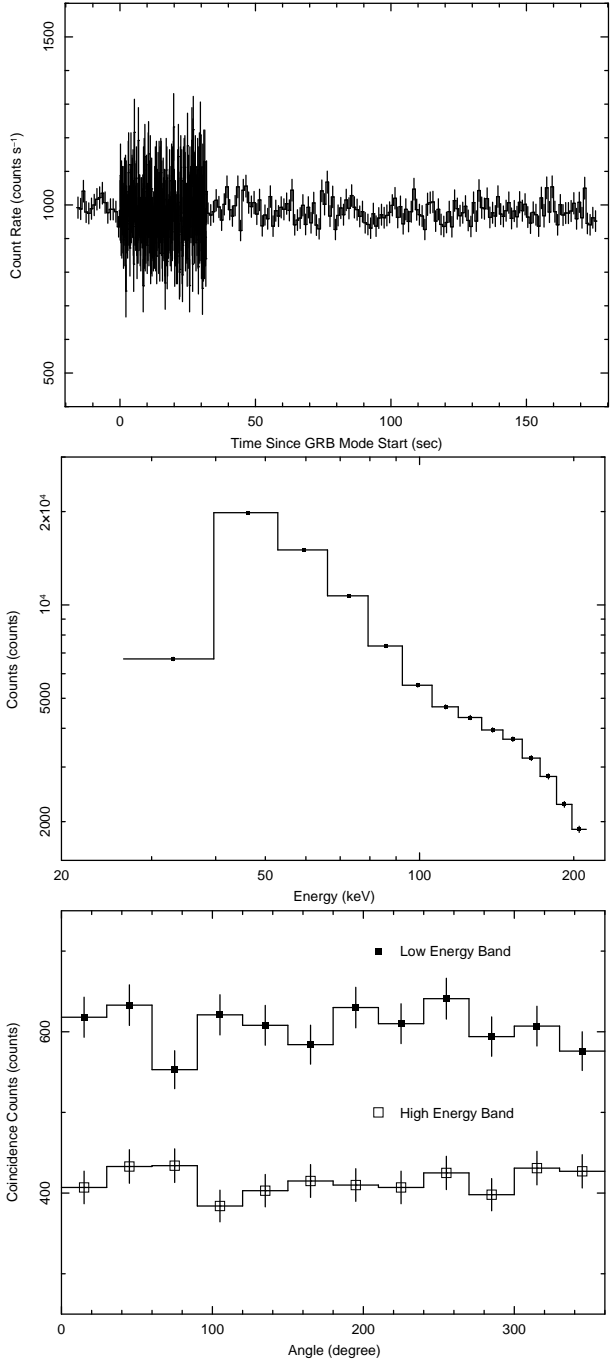

\rotatebox{270}{\includegraphics[width=60mm]{fig6a.ps}}\\
\rotatebox{270}{\includegraphics[width=60mm]{fig6b.ps}}\\
\rotatebox{270}{\includegraphics[width=60mm]{fig6c.ps}}
\caption{Data examples of the GRB mode, obtained by sending 
the manual GRB-mode command in the orbit. 
These data are equivalent to the cosmic X-ray/gamma-ray background.
(top) The lightcurve of gamma-ray events between LD and UD of
each sensor. The time resolution of the first 32~sec since trigger
time ($t=0$) is 125~msec, so the error bars accompanying 
the data becomes large.
(middle) The background spectrum summed of 12~CsI sensors between the
time interval of $t=0$ and $t=176$~sec in 16 PHA channels.
(bottom) The polarization data of the low- and high-energy band 
between the time interval of $t=-16$ and $t=176$.
}
\label{fig:grbmode}
\end{figure}

We can also start the GRB mode manually by sending 
a mode change command to get the background data in the manual GRB mode.
The top panel shows lightcurve, spectrum and polarization (coincidence) 
data of the GRB mode in figure~\ref{fig:grbmode}. We show the GRB 
lightcurve with the energy range between LD and UD, and it contains 
the data with two kinds of time resolution. In the first 32~second since 
the GRB trigger ($0 \le T < 32$), the lightcurve is recorded with the high 
time resolution of 125~msec, which is suitable for the interplanetary 
network (IPN) to determine the GRB direction together with other satellites.
In the following 144~sec ($32 \le T < 176$) and the 16~sec of pre-trigger 
($-16 \le T < 0$), the lighrcurves are recorded with 1~sec time resolution.
At the same time, one integrated spectrum is obtained between 
the time interval of $0 \le T < 176$ with 16 PHA channels as shown in 
figure~\ref{fig:grbmode} middle panel. 
This spectral data contains the photons of 12~CsI scintillators only 
with the single hit event. 
Figure~\ref{fig:grbmode} (bottom) is the summed polarization data of 
the low- and high-energy band as a function of phase angle,
between the time interval of $t=-16$ and $t=176$ in the manual GRB mode, 
respectively. There is no fake modulation seen in the angular distribution.
This is the strong advantage of the geometrical symmetry of GAP's 
configuration as shown in figure~\ref{fig:model}.

\begin{figure}
\rotatebox{0}{\includegraphics[width=100mm]{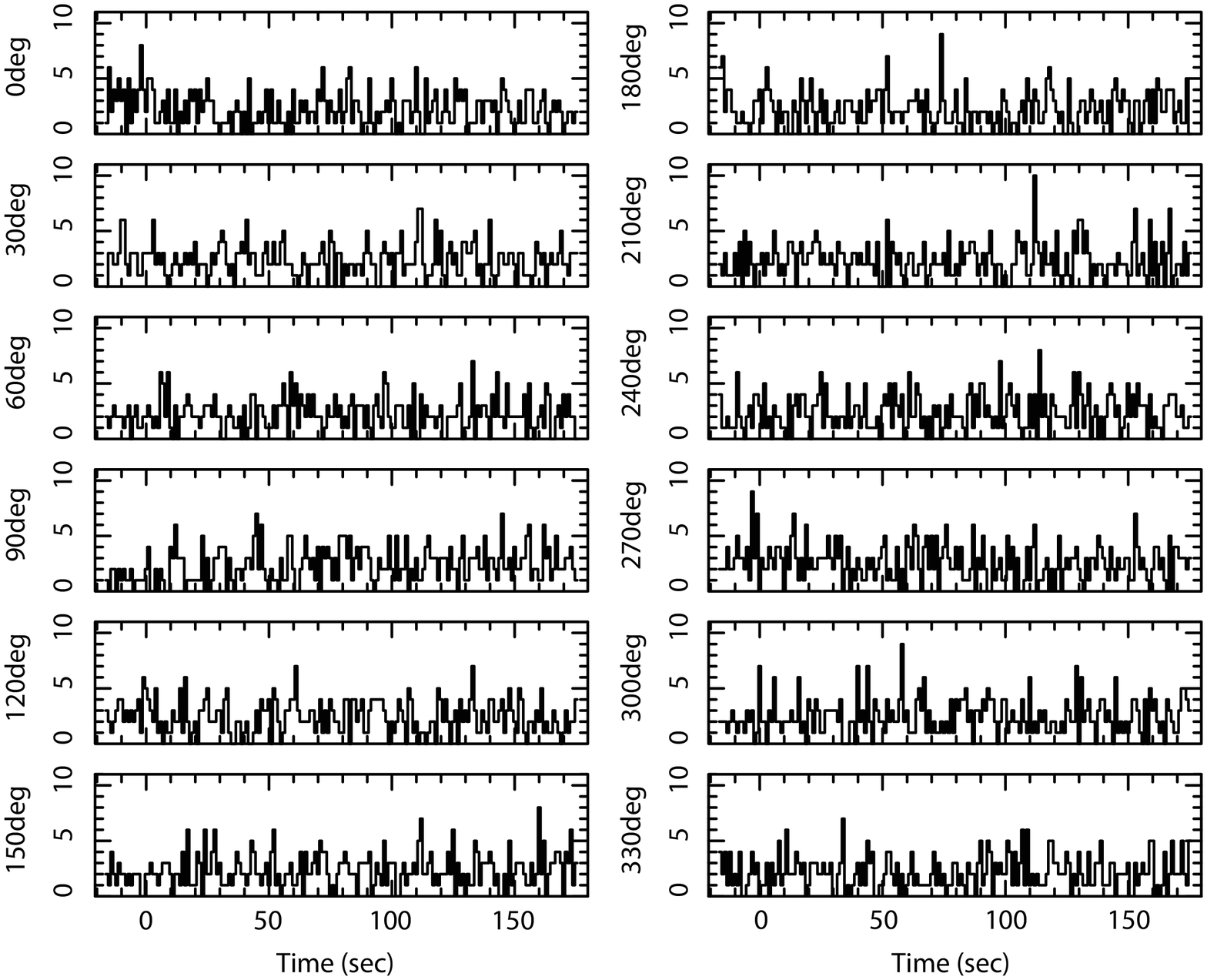}}
\caption{The polarization (hit-pattern) lightcurves with 
the time resolution of 1~sec obtained by the manual GRB mode. 
The time $t=0$~sec denotes the GRB trigger time. 
GAP always monitors the rotation period of IKAROS, 
so this lightcurve has been already corrected for
its rotation phase in every 125~msec. We can obtain these 
12~set of polarization data in two energy bands.
}
\label{fig:grbmode2}
\end{figure}

In figure~\ref{fig:grbmode2}, we show the lightcurves with 
the plastic--CsI coincidence events corrected for each angular phase
in the manual GRB mode. Here we determined the phase angle $\phi = 0$ 
as the CsI~No.1 when the sun pulse signal comes from the spacecraft.
When GAP enters the GRB mode, it starts measuring 
the polarization (hit-pattern of plastic--CsI) every one second. 
Therefore we are able to get the time resolved polarization data. 
GAP is always monitoring the rotation period of IKAROS, 
so this polarization lightcurve has been already corrected for 
its rotation phase in every 125~msec. The polarization is measured 
with 2~energy bands of High and Low. The energy range of the 
low-energy polarization is equivalent to the one between LD and 
HITPAT\_LH. Similarly the high-energy polarization is measured 
between HITPAT\_LH and UD. Here HITPAT\_LH is the boundary channel 
of energy dividing the low- and the high-energy, and we can 
change HITPAT\_LH with the command. The nominal value of HITPAT\_LH 
is set about $100~{\rm keV}$. 

When the GRB trigger is set, the GAP-CPU latches the GRB trigger 
time (clock counter), and store it in the SRAM. GAP also memorizes 
the sun pulse time (phase) around the trigger time and the rotation 
period of the IKAROS spacecraft at that time. Therefore we can 
determine the phase angle (meridian) of the polarization vector 
around the IKAROS rotation axis. 
GAP also stores total photon counts of each phase angle
(30~degree angular resolution) during the GRB mode, 
so we enable to determine very rough direction for bright GRBs.

\begin{table}
\caption{The time resolution of the GRB mode.}
\begin{tabular}{cccc}
\hline
time & lightcurve & polarization     & spectrum \\
    & (LD--UD)   & (low:LD--HITPAT\_LH) & (LD--UD)\\
    &            & (high:HITPAT\_LH--UD) & (16~channel)\\
\hline
$-16 \le T < 0$ & 1~sec    & 1~sec & N/A  \\
$0 \le T < 32$  & 125~msec & 1~sec & --- \\
$32 \le T < 176$    & 1~sec    & 1~sec & --- \\
\hline
$0 \le T < 176$ & ---      & ---   & 176~sec \\
\hline
\end{tabular}
\label{table-flux}
\end{table}

\subsection{CRAB Mode}

We prepared the mode, which can continuously observe targets, 
named CRAB mode. This is essentially the mode for continuous 
monitoring of the Crab. Using the phase information of 
IKAROS's rotation period measured with the sun pulse interval
(approximately 1~rpm), GAP can stack the polarization data 
during this Crab mode.
In Crab mode, GAP starts the integration of polarization data
at the time when the CPU gets the sun pulse signal. Therefore 
the phase angle of polarization vector can be determined.
Of course, when the count rate significantly increases and 
the GRB flag is happen during the CRAB mode, 
the CPU switches its status to the GRB mode.

According to the current plan of the IKAROS attitude,
GAP will observe along the ecliptic plane, which crosses at 
the galactic plane near the galactic center and also
at its opposite direction, very close to the Crab nebula. 
Thus we will spend 90~\% of the observable time in the CRAB mode 
continuously, and we try to measure the polarization from 
any hard X-ray and gamma-ray objects in the field of view. 
The Crab itself is expected to be in our field of view 
within 10~degree from the GAP's detector axis in 
the almost whole month of December, 2010. 
In this Crab mode, GAP obtains the spectrum of 
plastic scintillator, the summed spectrum of 12~CsI scintillator 
in 16 PHA channels, and polarization of 2~energy bands 
(LD--HITPAT\_LH and HITPAT\_LH--UD) with 24~angular resolution.
The exposure of each Crab mode can be set from 30~sec to 
8.5~hours, and we usually set the Crab mode with about 110~minutes
integration.

\begin{figure}
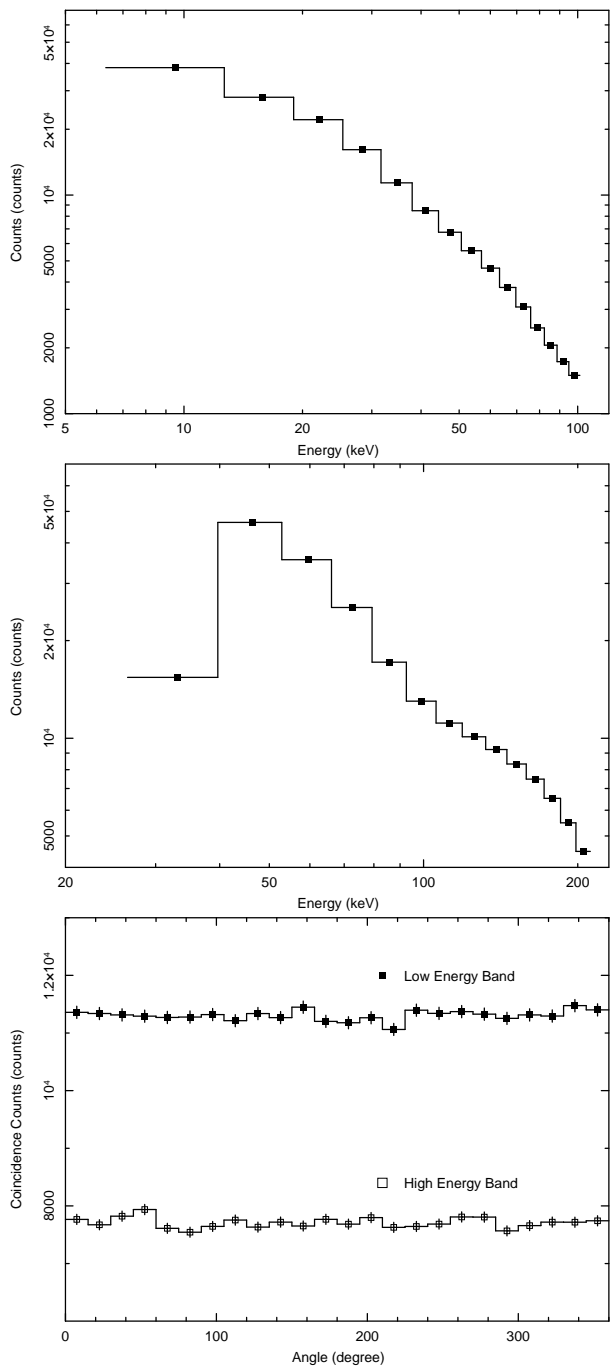

\rotatebox{270}{\includegraphics[width=60mm]{fig8a.ps}}\\
\rotatebox{270}{\includegraphics[width=60mm]{fig8b.ps}}\\
\rotatebox{270}{\includegraphics[width=60mm]{fig8c.ps}}
\caption{Data examples of the Crab mode during the 110~minutes
integration time. The Crab itself was out of FOV.
Instead, the Galactic center was in the FOV. Therefore 
the data are the cosmic X-ray/gamma-ray background and 
the sources near the Galactic center.
(top) The spectrum of the plastic scintillator.
(middle) The spectrum of 12~CsI sensors. 
(bottom) The polarization data of the low- and high-energy band.
}
\label{fig:crabmode}
\end{figure}

We show two spectra in the Crab mode for the plastic and 
CsI detectors during the 110~minutes integration 
in figure~\ref{fig:crabmode}. The Crab itself was out of 
the field of view. If GAP observes the sky without any strong 
X-ray/gamma-ray sources, 
these two spectra should be mainly explained by the Cosmic X-ray 
Background (CXB). The particle background was rejected by 
the UD of each scintillator and the multiple coincidence logic.
However the GAP has the wide field of view,
and observed the almost galactic center during the first 
two months since the GAP switch-on. Therefore these two spectra 
are the mixture of the CXB and the GC sources. 

As shown in the bottom of figure~\ref{fig:crabmode}, 
both modulation curves are flat. When we adopt the constant model, 
the fitting results of low- and high-energy band are 
$(1.130 \pm 0.004) \times 10^{4}$ with $\chi^2 = 16.0$ for
23~degrees of freedom and $(7.706 \pm 0.030) \times 10^{3}$ 
with $\chi^2 = 24.3$ for 23~degrees of freedom, respectively.
Here the error is 90~\% confidence level.
The observed modulation is quite flat within 1~\% level for 
the long integration of the CXB and GC sources in the Crab mode. 
This strongly supports the flatness of the GAP angular response 
for the non-polarized data.

\section{Detector Capability Estimated by {\it Geant~4} Simulation}

\begin{figure}
\rotatebox{0}{\includegraphics[width=80mm]{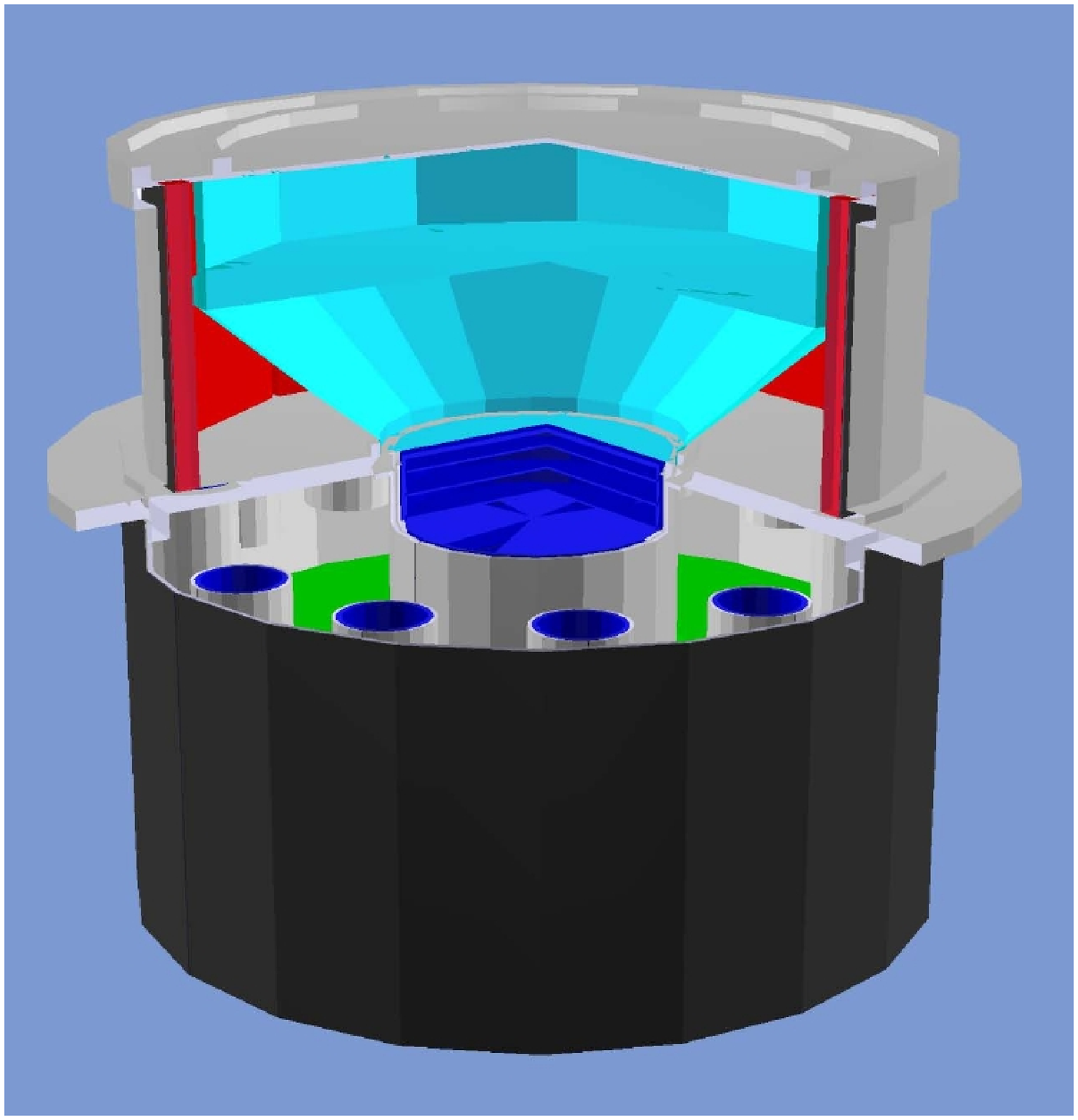}}
\caption{The geometrical mass model of GAP for Geant~4 simulator.
Scintillators (Plastic: water color and CsI: red), PMTs (blue), 
chassis case (Aluminum: grey, lead shields: dark grey), 
electronics circuits and  HV modules (green) with appropriate 
mass elements.
}
\label{fig:geant4mass}
\end{figure}

In this section, we show Monte Carlo simulations with the 
{\it Geant~4} simulator to estimate the capability of GAP and 
verify the model parameters. As shown in figure~\ref{fig:geant4mass}, 
we set a Geant-4 model of geometries and mass of elements which 
represents the real detector as precise as possible.
The cylindrical aluminum cases of detector (light gray) are covered 
by thin lead sheets (dark gray) except for the top panel window 
to avoid the scattered photons from the spacecraft body. 
So the direct X-ray from the top direction illuminates the plastic
and also the edge of the CsI. 
The plastic and 12~CsI scintillators are shown in light blue and red,
respectively. The 13 PMTs are shown in blue and all of them are covered 
with thin aluminum cylinders. The green sheet inside the detector case
represents the three layers of electric circuits. Two high voltage 
modules are also modeled while they are behind the detector case.
Additionally, we consider a simplified spacecraft body with 
thick aluminum sheets (mass equivalent) as a scatterer 
in the geometry model (not appeared in figure~\ref{fig:geant4mass}).
The outputs from this simulator are edited to the same type of 
informations observed by GAP. Moreover, to investigate the physical 
interaction between the incident gamma-ray photons and GAP, 
we also outputs higher order of informations such as spectrum of 
each sensor, coincidence spectrum of each sensor, 
polarization informations for single, double and multiple coincidence. 

\subsection{Modulation Factor,  Efficiency and pre-flight calibration }

\begin{figure}
\rotatebox{270}{\includegraphics[width=65mm]{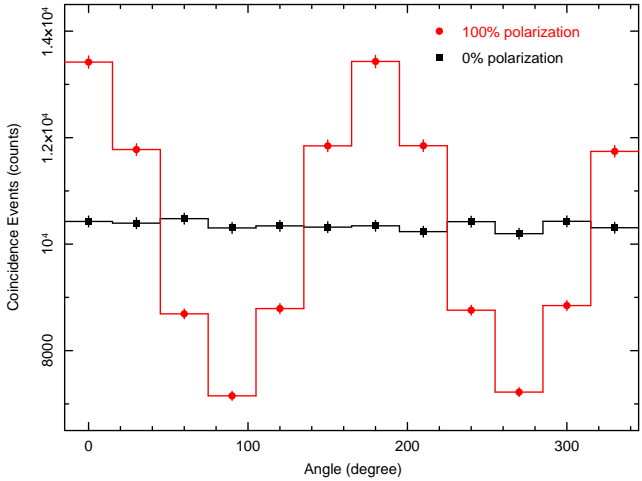}}\\
\caption{
Example of the modulation curve for monochromatic
gamma-ray of 100~keV with 100~\% polarization (filled circle) and 
non-polarization (filled square).
For the non-polarized gamma-ray, the modulation curve becomes 
constant and its deviation is under 1~\%. 
}
\label{fig:sim-modu}
\end{figure}

A value of modulation factor ($M_{100}$) is the key parameter 
to describe the detector sensitivity for the polarization 
measurement. The definition of the $M_{100}$ is the ratio of
the amplitude to the average level of modulation curve of 
detector response for 100~\% polarized gamma-ray. 
According to the cross section of the Compton effect in 
equation~(\ref{eq:klein-nishina}) and (\ref{eq:thomson}),
the angular distribution of the scattering photon basically shows
$\sin^{2} \phi$ modulation. 
Figure~\ref{fig:sim-modu} shows the simulated modulation
curves when we assume the condition that the gamma-ray photons 
with 100~\% polarization degree (open square) and non-polarization 
(filled square) are irradiated in front of GAP.
The GAP's modulation factor $M_{100}$ is estimated by the
Geant~4 simulator as
\begin{eqnarray}
\label{eq:modu}
M_{100} \equiv \frac{N_{max} - N_{min}}{N_{max} + N_{min}} 
= \frac{{\rm amplitude~of~modulation}}{{\rm average~of~modulation}}.
\end{eqnarray}
Here, $N_{max}$ and $N_{min}$ is the maximum and minimum counts 
of the angular distribution of scattering photons.

\subsection{Ground-base Calibration at KEK Photon Factory} 

To verify the mass model and also check the logics in 
the Geant-4 simulator, we carried out the ground-based calibration 
using the Proto-Flight Model (PFM-GAP) before launch.
The Flight Model (FM-GAP) was not ready to use at that time, 
so we used the PFM-GAP for the test on November 1--5, 2008 
at KEK Photon Factory (beam line PF-14A). 
The only difference between FM-GAP and PFM-GAP is the thickness 
of the central plastic and height of CsI. 
The PFM-GAP is 7~cm in thickness and height, but we change 
the design of FM-GAP into 6~cm to reduce the weight of the FM system. 

\begin{figure}
\rotatebox{270}{\includegraphics[width=65mm]{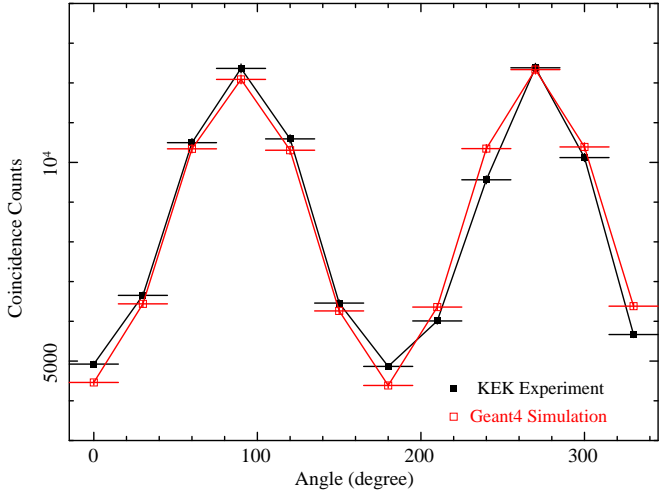}}
\caption{
The modulation curves for the monochromatic 80~keV with 82~\% polarized 
X-ray beam measured by the experiment at KEK Photon Factory (red points), 
and the result by Geant-4 simulation for the PFM-GAP model 
in the same condition.
Both of modulation curves show clear agreement with each other, 
and this fact strongly supports the validity of the mass model 
for PFM-GAP in Geant-4. The same mass model is used for the FM-GAP except 
the thickness of plastic.
}
\label{fig:KEK-Geant4}
\end{figure}

We tested a very limited number of parameters such as the modulation 
factor (MF) and also the detection efficiency ($\eta$). 
The monochromatic 80~keV pencil beam of 0.8~mm in diameter 
with 82~\% polarization was irradiated to the center of PFM. 
The beam is so narrow that we could not carry out the uniform 
irradiation on the top surface. The observed distributions of 
coincidence photons of the 12~CsI units are shown in 
figure~\ref{fig:KEK-Geant4} together with the Geant-4 simulation 
for the mass model.
The black and red points are the result of KEK experiment and 
that of Geant-4 simulation by the experiment beam, respectively. 
Both curves agree well and strongly support the validity of 
the Geant-4 mass model for the PFM-GAP detector. 
The modulation factor for the 82~\% polarized beam at the center 
position is $MF =0.43$ by the experiment. This is equivalent to 
$M_{100} = 0.52$ for the 100~\% polarized beam irradiated to 
the center of PFM-GAP.

\begin{figure}
\rotatebox{270}{\includegraphics[width=65mm]{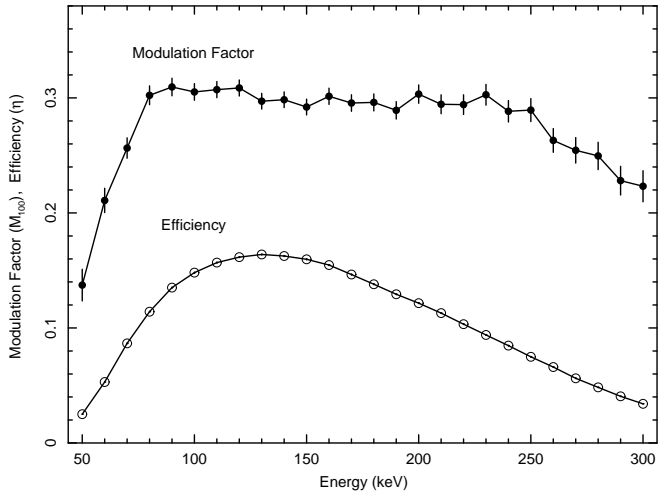}}
\caption{
The modulation factor and the efficiency of FM-GAP as a function of 
incoming gamma-ray energy calculated by Geant~4 simulation. 
We determine the modulation factor with equation~(\ref{eq:modu}). 
The sensitivity of GAP is optimized around the energy of 80--150~keV.
}
\label{fig:m-eta}
\end{figure}

In figure~\ref{fig:m-eta}, we show the modulation factor and 
an efficiency, which are calculated for the mass model of FM-GAP 
using the same logics of the PFM, as a function of photon energy 
in case of uniformly irradiated into the full surface of FM-GAP. 
The modulation factor for the FM-GAP is obtained as $M_{100} = 0.3$ 
between 80--200~keV range, using the same mass model except for 
the thickness of the plastic and the height of CsI.
In case of uniform irradiation on the top surface, 
this value is lower than the KEK pencil beam experiment. 
The efficiency ($\eta$), which  is defined as a ratio of 
the number of plastic--CsI coincidence events to the total 
incoming photons is very low of about 0.18 at around 
100--150~keV. Since the depth and diameter of the plastic 
scintillator is optimized as 1~Compton length for 100~keV photon, 
the self-absorption reduces the efficiency and the modulation 
factor for the lower energy photon below 100~keV.
The value of modulation factor at low-energy is also strongly 
influenced by the setting of LD ($\sim 7~{\rm keV}$) of 
plastic scintillator.

\subsection{Minimum Detectable Polarization}
Using the modulation factor ($M_{100}$) and the efficiency 
shown in figure~\ref{fig:m-eta}, we estimate a rate of GRBs
whose polarization is detected. We use the definition of 
minimum detectable polarization (MDP) as the sensitivity for 
the polarization measurement (e.g. \cite{weisskopf2010}). 
The MDP is defined as
\begin{eqnarray}
\label{eq:mdp}
MDP &=&
\frac{n_{\sigma}}{M_{100}} \frac{\sqrt{(F S \eta + B) T}}{F S \eta T}.
\end{eqnarray}
\begin{eqnarray*}
F &:& \mbox{GRB photon flux}~({\rm counts~cm^{-2}~s^{-1}}),\\
B &:& \mbox{background count rate}~({\rm counts~s^{-1}}),\\
S &:& \mbox{geometrical area}~({\rm cm^{2}}),\\
T &:& \mbox{GRB duration}~({\rm sec}),
\end{eqnarray*}
Here, the $n_{\sigma}$ value is the significance of detection.
If we assume 99~\% confidence level, $n_{\sigma} = 3 \sqrt{2}$ 
is generally used \citep{weisskopf2010}. .

The detectability of polarization depends on the source flux ($F$), 
the total duration ($T$) and the area of detector ($S$) for a GRB, 
that is to say a total number of photons. To estimate the MDP, 
we used a model of the brightness and duration distribution of 
CGRO-BATSE catalog of 50--300~keV range which is almost matching 
the energy coverage of GAP. If we assume a case that the polarization 
degree of GRBs shows about $40$~\% level \citep{lazzati2004, toma2009}, 
we expect to detect the polarization with a rate of 
2--4~GRBs/year above the 40~\% degree of polarization.
If a degree of polarization is lower than this, a chance of 
detection will be one or two a year.

\subsection{Response for Polarized Beam of Diagonal Incident Photon}
\begin{figure}
\rotatebox{270}{\includegraphics[width=80mm]{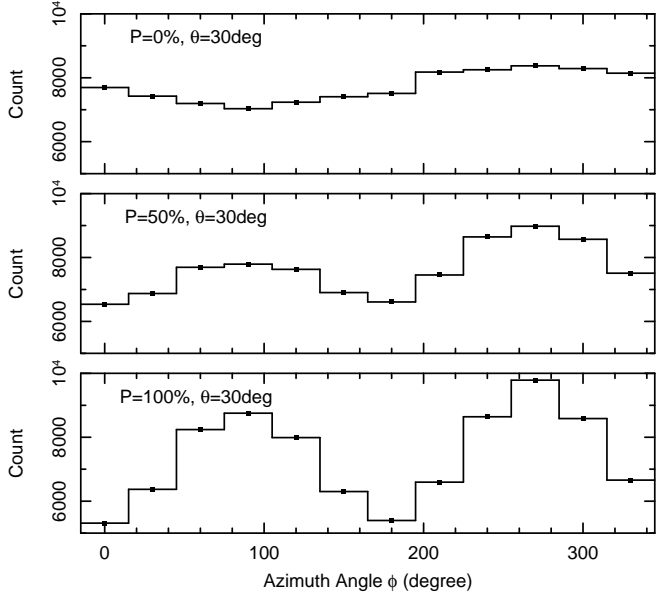}}
\caption{
Examples of simulated modulation curve by {\it Geant 4} for 
the angle of incident photon as $\theta = 30$~deg,
and the azimuth angle of $\phi = 90$~deg. 
The top, middle and bottom panels show the modulation curve 
of polarization degree of $P = 0, 50, 100$~\%, respectively, 
in case of the power-law spectrum with the photon index of $-1$. 
We assume the polarization vector is parallel to the detector 
surface in these simulations.
}
\label{fig:30deg}
\end{figure}

In figure~\ref{fig:30deg}, we show the detector response 
simulated by the Geant-4 for the incident photons from 
the diagonal direction. We assume the incident direction
of $\theta = 30$~deg from the detector axis of GAP, 
and the azimuth angle of $\phi = 90$~deg. 
The top, middle and bottom panels show the response 
of polarization degree of $P = 0, 50, 100$~\%, respectively.
We assume the polarization vector is parallel to the detector 
surface in these simulations.

In the case of $P = 0$~\%, the modulation curve basically depends
on the $\sin \phi$. This is because the GAP surface looks like 
elliptic shape from the diagonal position. 
On the other hand, the modulation curve of polarized gamma-ray 
shows the composition of $\sin^2 \phi$ (see equation~(\ref{eq:compton}))
and $\sin \phi$ by the response for the diagonal incident photons.
As shown in the middle and bottom panels of figure~\ref{fig:30deg}, 
the modulation with two peaks will be observed for polarized gamma-rays, 
but the peak intensity is different from each other by
the diagonal effect. Not to say, the position of the peak is 
also influenced by the vector of incident polarization.

\section{First Detection of GRB}
\begin{figure}
\rotatebox{270}{\includegraphics[width=65mm]{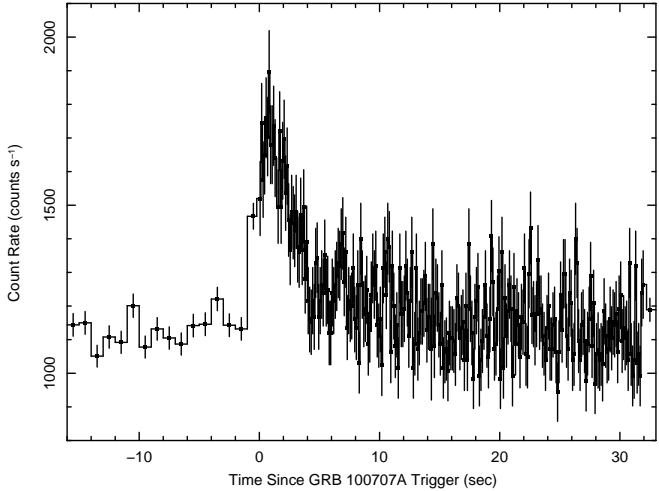}}
\caption{The lightcurve of GRB~100707A detected by IKAROS-GAP.
This event occurs at $\theta = 90$~degree measured from 
the GAP detector axis. 
}
\label{fig:grb100707a}
\end{figure}

Since the switch-on of June 21, GAP detected the first GRB of 
GRB~100707A on July 7, 2010. The lightcurve is shown in
figure~\ref{fig:grb100707a}. The trigger time is $T_{0}=$~00:47:22.878~UT
with the uncertainty of $\pm 0.250$~sec at the distance of 17,917,783~km 
from the Earth. The direction of IKAROS from the Earth on July 7 is 
$(\alpha, \delta) = (-150.29, -24.32)$. 
The time accuracy will be improved to $\pm 0.125$~sec in near future 
after the careful calibration of clock system.
This burst was also triggered by Fermi-GBM, Suzaku-WAM, Konus-Wind,
and Messenger-GRN. Therefore the direction of GRB~100707A was
determined  by the interplanetary network (IPN). 
The error box of GRB direction by IPN is reported as
$(\alpha, \delta) = (358.019^{+0.754}_{-0.903}, -8.658^{+3.363}_{-2.996})$
by \citet{ipn}.

We have checked the consistency of our clock system to the reported 
IPN GRB direction. Cross-correlating the lightcurves between 
the IKAROS-GAP and Suzaku-WAM, the difference of arrival time of 
maximum photon flux between GAP and Suzaku-WAM is 
$\Delta T = 42.230 \pm 0.250$~sec. Then the GRB position is calculated 
as the corn with the angle of $\theta_{\rm GAP} = 43.68 \pm 0.36$~degree 
from the direction of IKAROS. Here we assumed the error of angle is 
due to the time accuracy (0.250~sec) of GAP system. 
The estimated GRB direction is $\theta_{\rm IPN} = 45.21^{+1.77}_{-1.80}$
in 3~$\sigma$ uncertainty which is consistent with $\theta_{\rm GAP}$.
Therefore we conclude that GAP can join the IPN system, 
which consists of such as Fermi, Konus-Wind, Messenger-GRN and so on.

According to the IKAROS attitude, the axis of GAP direction is 
$(\alpha, \delta) = (265.257, -18.337)$ on July 7, 2010. 
Therefore, the incident direction of GRB~100707A is 
almost $\theta = 90$~degree from the center of GAP field of view.
In this case, the polarization measurement is quite difficult.
According to the Konus-Wind results, the peak energy is 
$E_{p} = 264^{+49}_{-40}~{\rm keV}$ \citep{gcn10948},
and Fermi-LAT also detected this event \citep{gcn10945}, 
so the GRB~100707A was a hard and bright burst. 
This hard spectrum could enable GAP to detect this event 
even if the side cylinder and the bottom is shielded by the thin lead.

\section{Summaries}
\begin{enumerate}
\item
We developed the first gamma-ray burst polarimeter ``GAP'' aboard 
the small solar power sail demonstrator IKAROS. 
\item
The IKAROS spacecraft was launched on May 21, 2010. 
We performed calibrations since the switch-on of June 21, 2010, 
and succeeded in setting all CsI gains within 2~\%. 
After that we started GRB observation. 
\item
The geometrical shape of GAP is highly symmetry, so the systematic 
uncertainty of fake modulation is about 1~\% level for the CXB.
The pre-flight calibration at KEK and the Geant-4 simulation, 
which is fully consistent, strongly support the validity of our 
detector. GAP measures the angular anisotropy of Compton scattering
with real coincidence events between the plastic and 12~CsI 
scintillators. These two points are advantage for the gamma-ray 
polarimetry, and we will perform reliable observation for 
the gamma-ray polarization of GRBs, bright magnetar flares, and
Crab nebula.
\item
According to the current operation, the detection rate of GRBs is
about once a week. For the bright events, we can discuss 
the polarization degree. The rate of expectation to detect 
the polarization is 2--4 events a year if GRBs show about 40~\% 
polarization as theoretical prediction.
\item
GAP will observes the gamma-ray polarization in the entire field 
of view with the Crab mode. We will measure the polarization degree 
or upper-limit along to the ecliptic plane, especially for 
the Crab nebula and the galactic center.
\item
We require informations about the GRB direction while 
we will contribute to the interplanetary network system soon.
\end{enumerate}

\section*{Acknowledgments}
We deeply thank the operations staffs at the Tokyo Metropolitan 
Industrial Technology Research Institute and 
Dr. Satoshi HATORI, Dr. Kyo KUME and technical staff at 
the Wakasa-Wan Energy Research Center for their assistances 
during the qualification tests of electrical devices. 
We also thank Mr. Takano at the Industrial Research Institute of 
Ishikawa for his support during the mechanical test. 
This work is supported in part by the Grant-in-Aid from the 
Ministry of Education, Culture, Sports, Science and Technology
(MEXT) of Japan, No.20674002 (DY), No.18684007 (DY) 
and also this is supported by the Steering Committee for 
Space Science at ISAS/JAXA of Japan (TM).


%

\begin{thebibliography}{99}
\bibitem[Amati et al.(2002)]{amati02}
Amati, L., Frontera, F., Tavani, M, et al. 2002, A\&A, 390, 81
%
\bibitem[Amati et al.(2006)]{amati06}
Amati,~L., 2006, MNRAS, 372, 233
%
\bibitem[Bellazzini et al.(2003)]{bellazzini}
Bellazzini, R., Angelini, F., Baldini, L., et al., 2003, SPIE, 4843, 383
%
\bibitem[Coburn \& Boggs(2003)]{coburn}
Coburn, W. \& Boggs, S. E., 2003, Nature, 423, 415
%
\bibitem[Costa et al.(1997)]{costa}
Costa, E., Frontera, F., Heise, J., et al., 1997, Nature, 387, 783
%
\bibitem[Dean et al.(2008)]{dean2008}
Dean, A. J., et al., 2008, Science, 321, 1183
%
\bibitem[Fenimore \& Ramirez-Ruiz(2000)]{fenimore}
Fenimore, E. E., \& Ramirez-Ruiz, E. \ 2000, astro-ph/0004176
%
\bibitem[Ghirlanda et al.(2004)]{ggl04}
Ghirlanda,~G., Ghisellini,~G.,  Lazzati,~D., 2004, ApJ, 616, 331
%
\bibitem[Golenetskii et al.(2010)]{gcn10948}
Golenetskii, S., Aptekar, R., Frederiks, D., et al., 2010, GCN, 10948
%
\bibitem[Gunji et al.(1994)]{gunji1994}
Gunji, S., Sakurai, H., Noma, M., et al., 1994, IEEE, 41, No.4
%
\bibitem[Gunji et al.(1997)]{gunji1997}
Gunji, S., Kudo, E., \& Sakurai, H., 1997, IEEE, 41, No.4
%
\bibitem[Gunji et al.(2007)]{gunji2007}
Gunji, S., et al., 2007, SPIE, 6686, 18-1 
%
\bibitem[Hamamatsu Photonics]{hamamatsu}
http://jp.hamamatsu.com/en/index.html
%
\bibitem[Hill et al.(2007)]{hill2007}
Hill, Joanne E.; Barthelmy, Scott; Black, J. Kevin, et al.,
2007, SPIE, 6686, 66860Y
%
\bibitem[Hurley et al.(2010)]{ipn}
Hurley, K., Goldsten, J., 2010, GCN, 10947
%
\bibitem[Kamae et al.(2008)]{kamae2008}
Kamae, T., Andersson, V., Arimoto, M., et al., 2008, 
Astroparticle Physics, 30, 72
%
\bibitem[Kawaguchi et al.(2008)]{kawaguchi2008}
Kawaguchi, J., Mori, O., Tsuda, Y., et al., 2008, 
Proceedings of 59th International Astronautical Congress, 
IAC-08.A3.6.15
%
\bibitem[Kestenbaum et al.(1976)]{kestenbaum1976}
Kestenbaum, H. L., Cohen, G. G., Long, K. S., et al., 1976, ApJ, 210, 805
%
\bibitem[Kishimoto et al.(2007)]{kishimoto07}
Kishimoto, Y.; Gunji, S.; Ishigaki, Y., et al., 2007,
IEEE Trans. on Nucl. Sci. 54, 3, 561--566
%
\bibitem[Kitamoto et al.(2010)]{kitamoto}
Kitamoto, S., Murakami, H., Shishido, Y., et al., 2010,
Review of Scientific Instruments, 81, 023105
%
\bibitem[Lamb et al.(2004)]{lamb04}
Lamb,~D.~Q. et al.  2004, New Astron. Rev. 48, 423 (astro-ph/0309462)
%
\bibitem[Lazzati et al.(2004)]{lazzati2004} 
Lazzati, D., Rossi, E., Ghisellini, G, Rees, M. J., 2004, MNRAS, 347, L1
%
\bibitem[McGlynn et al.(2007)]{mcglynn}
McGlynn, S., Clark, D. J., Dean, A. J., et al.,
2007, A \& A, 446, 3, 895-904
%
\bibitem[Mihara \& Miyamoto(2004)]{mihara2004}
Mihara, T., \& Miyamoto, H., 2004, Proceedings of X-ray Polarimetry Workshop.
%
\bibitem[Mori et al.(2009)]{mori2009}
Mori, O., Sawada, H., Hanaoka, F., et al., 2009,
Transactions of the Japan Society for Aeronautical and Space Sciences, 
Space Technology Japan, Vol.7, No.ists26, pp.Pd87-Pd94
%
\bibitem[Murakami et al.(2010)]{murakami2010}
Murakami, T., Yonetoku, D., Gunji, S., et al., 2010, Proceedings of
Gamma-Ray Burst Conference in Kyoto
%
\bibitem[Norris et al.(2000)]{norris}
Norris,J., Marani, G.,\& Bonnell, J.  2000, ApJ, 534, 248
%
\bibitem[Rees \& Meszaros(1992)]{rees1992} 
Rees M. J. \& M\'{e}sz\'{a}ros P., 1992, MNRAS, 258, L41
%
\bibitem[Rutledge \& Fox(2004)]{rutledge04}
Rutledge, R. E.\& Fox, D. B., 2004, MNRAS, 350, 4, 1288-1300
%
\bibitem[Pelassa et al.(2010)]{gcn10945} 
Pelassa, V., Pesce-Rollins, M., et al., 2010, GCN, 10945
%
\bibitem[Piran(1999)]{piran1999} 
Piran, T., 1999, Physics Report, 314, 575
%
\bibitem[Sakamoto et al.(2004)]{sakamoto04}
Sakamoto, T., Lamb, D. Q., Graziani, C., et al. 2004, ApJ, 602, 875
%
\bibitem[Salvaterra et al.(2009)]{salvaterra2009}
Salvaterra, R. et al., 2009, Nature 461, 1258-1260
%
\bibitem[Schaefer et al.(2001)]{schaefer01}
Schaefer, B. E., Deng, M. \& Band, D. L.  2001, ApJ, 563, L123
%
\bibitem[Suzuki et al.(2006)]{suzuki2006} 
Suzuki, T., Gunji, S., Nakajima, R., et al., 2006, 
Japanese Journal of Applied Physics, 45, 1
%
\bibitem[Swank et al.(2008)]{swank} 
Swank, J., Kallman, T.,; Jahoda, K.,
37th COSPAR Scientific Assembly. Held 13-20 July 2008, 
in Montr al, Canada., p.3102
%
\bibitem[Tamagawa et al.(2006)]{tamagawa} 
Tamagawa, T., Hayato, A., Yamaguchi, Y., et al., 2006,
SPIE, 6266, 62663W
%
\bibitem[Tanimori et al.(2004)]{tanimori2004} 
Tanimori, T., Kubo, H., Miuchi, K., et al., 2004, New Astron.Rev. 48, 263--268
%
\bibitem[Tanvir et al.(2009)]{tanvir2009} 
Tanvir, N. R. et al., 2009, Nature, 461, 1254--1257
%
\bibitem[Toizumi et al.(2009)]{toizumi2009} 
Toizumi, T., Nakamori, T., Kataoka, J., et al., 2009,
AIP Conference Proceedings, Volume 1133, 85-87
%
\bibitem[Toma et al.(2009)]{toma2009} 
Toma, K., Sakamoto, T., Zhang, B., et al., 2009, ApJ, 698, 1042
%
\bibitem[Weisskopf et al.(1976)]{weisskopf1976}
Weisskopf, M. C., Cohen, G. G., Kestenbaum, H. L., et al., 1976,
ApJ, 208, L125--L128
%
\bibitem[Weisskopf et al.(2010)]{weisskopf2010}
Weisskopf, M. C., Elsner, R. F., and O'Dell, S. L., 2010,
arXiv:1006.3711v2
%
\bibitem[Wigger et al.(2004)]{wigger04}
Wigger, C., Hajdas, W., Arzner, K. et al., 2004, ApJ, 613, 1088-1100
%
\bibitem[Yonetoku et al.(2004)]{yonetoku04}
Yonetoku, D., Murakami, T., Nakamura, T., et al. 2004, ApJ, 609, 935
%
\bibitem[Yonetoku et al.(2006)]{yonetoku2006}
Yonetoku, D., Murakami, T., Masui, H., et al., 2006, SPIE, 6266, 86
%
\bibitem[Yonetoku et al.(2009)]{yonetoku2009}
Yonetoku, D., Murakami, T., Gunji, S., et al., 2009, Proceedings of
X-ray Polarimetory in Rome
%
\end{thebibliography}
\end{document}